\documentclass{aa}  

\usepackage{graphicx}
\usepackage{txfonts}
\def\hi{\ifmmode {\mbox H{\scshape i}}\else H{\scshape i}\fi\xspace}
\def\hii{\ifmmode {\mbox H{\scshape ii}}\else H{\scshape ii}\fi\xspace}
\def\h2{\ifmmode {\mbox H$_2$}\else H$_2$\fi\xspace}
\def\micron{\ifmmode {\mbox $\mu$m}\else $\mu$m\fi\xspace}
\newcommand{\um}{$\mu$m}

\newcommand{\dtg}{\ensuremath{{\rm DTG}}}

\begin{document}

   \title{The dust-to-gas mass ratio of luminous galaxies as a function of their metallicity at cosmic noon}
    \titlerunning{The DTG of z$=$2.1--2.5 star-forming main-sequence galaxies as a function of metallicity}
    \authorrunning{G. Popping et al.}

   \author{Gerg\"o Popping,
          \inst{1}
          Irene Shivaei,\inst{2}
          Ryan L. Sanders,\inst{3} 
          Tucker Jones,\inst{3}
          Alexandra Pope,\inst{4}
          Naveen A. Reddy,\inst{5}
          Alice E. Shapley,\inst{6}
          Alison L. Coil,\inst{7}
          \and 
          Mariska Kriek\inst{8}
          }

   \institute{European Southern Observatory, Karl-Schwarzschild-Str. 2, D-85748, Garching, Germany\\
              \email{gpopping@eso.org}
         \and Steward Observatory, University of Arizona, Tucson, AZ
85721, USA
         \and Department of Physics, University of California, Davis, One
Shields Ave, Davis, CA 95616, USA 
        \and Department of Astronomy, University of Massachusetts, Amherst, MA 01003, USA
        \and Department of Physics and Astronomy, University of California, Riverside, 900 University Avenue, Riverside, CA 92521, USA
        \and Department of Physics \& Astronomy, University of California, Los Angeles, 430 Portola Plaza, Los Angeles, CA 90095, USA
        \and Center for Astrophysics and Space Sciences, University of California, San Diego, 9500 Gilman Dr., La Jolla, CA 92093-0424, USA
        \and Leiden Observatory, Leiden University, P.O. Box 9513, NL-2300 AA Leiden, The Netherlands
        }


 
  \abstract
   {}
   {We aim to quantify the relation between the dust-to-gas mass ratio (\dtg) and gas-phase metallicity of $z=$2.1--2.5 luminous galaxies and contrast this high-redshift relation against analogous constraints at z$=$0.}
   {We present a sample of ten star-forming main-sequence galaxies in the redshift range $2.1<z<2.5$   with rest-optical emission-line information available from the MOSDEF survey and with ALMA 1.2 millimetre and CO J$=$3--2 follow-up observations. The galaxies have stellar masses ranging from  $10^{10.3}$ to $10^{10.6}\,\rm{M}_\odot$ and cover a range in star-formation rate from 35 to 145 $\rm{M}_\odot\,\rm{yr}^{-1}$. We calculated the gas-phase oxygen abundance of these galaxies from rest-optical nebular emission lines (8.4 < $12 + \log{(\rm{O/H})} < 8.8$, corresponding to 0.5 -- 1.25 Z$_\odot$). We estimated the dust and \h2 masses of the galaxies (using a metallicity-dependent CO--to--\h2 conversion factor)  from the 1.2~mm and CO J$=$3--2 observations, respectively, from which we estimated a \dtg. }
   {We find that the galaxies in this sample follow the trends already observed  between CO line luminosity and dust-continuum luminosity from $z=0$ to $z=3$, extending such trends to fainter galaxies at $2.1<z<2.5$ than observed to date. We find no second-order metallicity dependence in the CO -- dust-continuum luminosity relation for the galaxies presented in this work. The \dtg s\ of main-sequence galaxies at $2.1<z<2.5$ are consistent with an increase in the \dtg\ with gas-phase metallicity. The metallicity dependence of the \dtg\ is driven by the metallicity dependence of the CO--to--\h2 conversion factor. Galaxies at z$=$2.1--2.5 are furthermore consistent with the DTG-metallicity relation found at z$=$0 (i.e. with no significant evolution), providing relevant constraints for galaxy formation models. These results furthermore imply that the metallicity of galaxies should be taken into account when estimating cold-gas masses from dust-continuum emission, which is especially relevant when studying metal-poor low-mass or high-redshift galaxies.  }
   {}


   \maketitle
%

\section{Introduction}
Dust is one of the key ingredients in interstellar medium (ISM) and galaxy physics. Dust influences interstellar chemistry (e.g. \citealt{Hollenbach1971}) and acts as a catalyst for the formation of molecular hydrogen, the fuel for star formation (e.g. \citealt{Cazaux2009}). Dust depletes metals from the gas-phase ISM \citep{Calzetti1994}, contributes to the metals in the circumgalactic medium \citep{menard2010}, and can even act as an additional cooling channel of hot gas onto galaxies \citep{Ostriker1973}. Dust grains absorb optical and ultraviolet stellar radiation and re-emit them at infrared wavelengths \citep{Spitzer1978}. Our understanding of star formation in the Universe is highly incomplete unless we account for the dust-obscured emission from young stars \citep[e.g.][]{Madau2014}.

The dust-to-gas mass ratio (\dtg ) of galaxies quantifies the fraction of ISM mass locked up in dust grains. The relation between gas-phase metallicity and the \dtg\  provides meaningful constraints on dust growth and destruction in galaxies and the timescale for star formation \citep{Asano2013, Zhukovska2014}. A number of semi-analytical \citep{Popping2017, Vijayan2019, Triani2020, Dayal2022} and hydrodynamic \citep{McKinnon2018, Li2019, Hou2019,Graziani2020} models of galaxy formation and evolution have included the production and destruction of dust. Predictions for the functional shape and time evolution of the relation between metallicity and the \dtg\ vary significantly between these models, with differences as high as an order of magnitude at fixed metallicity (see \citealt{Popping2022}). An observational census of the relation between gas-phase metallicity and the \dtg\ over cosmic time is thus timely. 

In the last two decades, a number of studies have demonstrated that the \dtg\ of local luminous galaxies increases as a function of gas-phase metallicity (e.g. \citealt{Remy-Ruyer2014}, \citealt{DeVis2019}, \citealt{Galliano2021}).  Absorption studies of damped Lyman-$\alpha$ absorbers have provided a statistical view of the relation between metallicity and the \dtg\ at $z>0$ (e.g. \citealt{DeCia2016,Wiseman2017, Peroux2020}), indicating that this relation is constant from $z=0$ to $z=5$ \citep{Popping2022}. By nature, these studies probe the \dtg\ of a galaxy along an individual sightline in the neutral ISM, rather than measuring the integrated \dtg, and thus do not probe the majority of the metals and dust within galaxies. 

Using a combination of Atacama Large Millimeter/submillimeter Array (ALMA) and MOSFIRE observations of four luminous $z>1$ main-sequence galaxies, \citet{Shapley2020} find that the normalisation of the relation between gas-phase metallicity and the \dtg\ appears constant with time out to $z\sim 2.5$.  However, this study was limited to approximately solar-metallicity objects. Missing from our current observational picture is thus a sample of luminous $z>0$ galaxies that covers a wide range in gas-phase metallicities for which integrated \dtg\ estimates are available. Such a sample will provide new constraints for theoretical models on the functional shape and evolution of the relation between gas-phase metallicity and the \dtg. 

In recent years, the dust-continuum emission of galaxies has frequently been used as a tracer of cold-gas mass \citep[e.g.][]{Eales2012, Groves2015, scoville16, Schinnerer2016, Kaasinen2019}. At z$>$0 these relations were calibrated against bright and metal-rich sub-millimetre galaxies \citep{scoville16}. The validity of this methodology for measuring cold-gas mass in sub-solar-metallicity $z>0$ galaxies has not yet been probed. A complete observational census of the functional shape and evolution of the relation between the \dtg\ and gas-phase metallicity can thus provide new insights into the use of dust-continuum emission as a cold-gas mass tracer.

In this paper we present the relation between gas-phase metallicity and the \dtg\ for ten $z\sim 2.3$ main-sequence galaxies. These galaxies have robust metallicity estimates available from the MOSFIRE Deep Evolution Field (MOSDEF) survey \citep{Kriek2015} and dust and \h2 mass constraints from ALMA observations \citep{Shivaei2022, Sanders2022}. In Sect.~\ref{sec:data} we present the sample and the MOSFIRE and ALMA data, and we discuss how galaxy metallicity and the \dtg\ were derived. Section~\ref{sec:results} presents the results of this study, focusing on the relation between CO and dust continuum luminosity and the metallicity--\dtg\ relation. We discuss and summarise our results in Sect.~\ref{sec:discussion}. We adopt a $\Lambda$ cold dark matter cosmology with $H_0 = 70$\,km\,s$^{-1}$\,Mpc$^{-1}$, $\Omega_{\Lambda}=0.7$, and $\Omega_m=0.3$ and a \citet{Chabrier2003} initial mass function, {and we assume a solar metallicity of 12 + log(O/H) = 8.69 \citep{Asplund2009}.}

\begin{figure}
\centering
\includegraphics[width=0.9\linewidth]{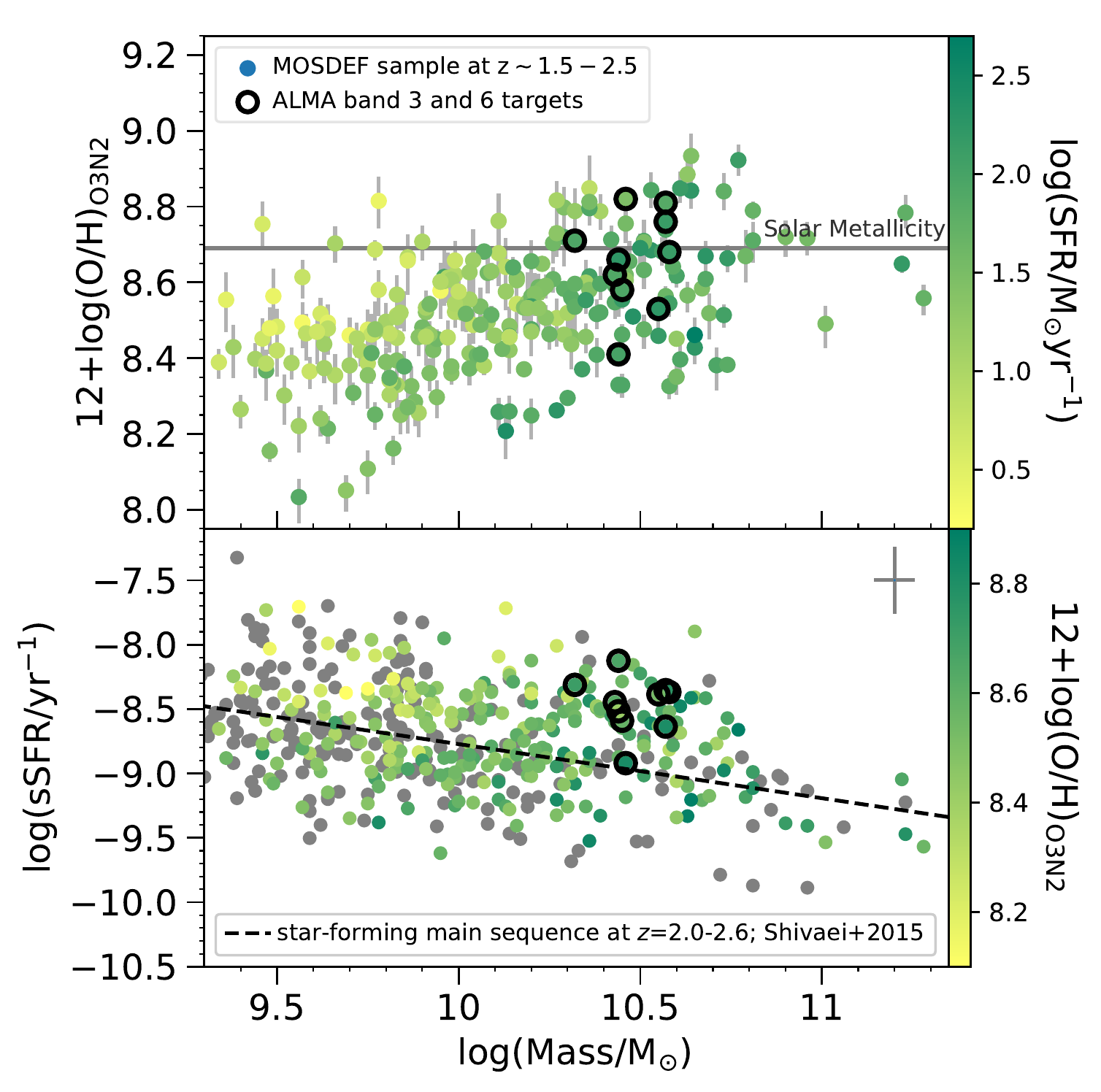}
\caption{Metallicity, stellar mass, and specific SFR distributions of the ALMA targets (black open circles) compared to the parent MOSDEF sample at $z=$1.5--2.5. Solar metallicity is shown with the grey line in the upper panel. The star-forming main sequence at z$=$2.0--2.6 \citep{Shivaei2015} is shown by the dashed line in the lower panel.  \label{fig:MS} }
\end{figure}

\begin{table*}

\caption{Sample properties.\label{tab:sample}} 
\begin{tabular}{p{2.5cm}cccccc}
\hline
\hline
ID & Redshift & 12 + log(O/H) & $S_{1.2\,\rm{mm}}/\mu\rm{Jy}$ & $\log{(L'_{\rm CO3-2}/ [\rm{K}\,\rm{km}\,s^{-1}\,\rm{pc}^{2}])}$  & $\log{(M_{\rm{H2}}/\rm{M}_\odot)}$ & $\log{(M_{\rm dust}/\rm{M}_\odot)}$   \\      
\hline
19985&2.19&8.53$\,\pm\,$0.01&107$\,\pm\,$34&9.75$\,\pm\,$0.16&10.91$\,\pm\,$0.16&8.31$\,^{0.32}_{0.18}$\\
13701&2.17&8.76$\,\pm\,$0.02&$<$144&9.57$\,\pm\,$0.1&10.36$\,\pm\,$0.1&$<$8.44\\
5814&2.13&8.68$\,\pm\,$0.03&156$\,\pm\,$66&<9.86&<10.52&8.47$\,^{0.35}_{0.21}$\\
3666&2.09&8.41$\,\pm\,$0.02&$<$78&<9.8&<10.9&$<$8.18\\
9971&2.41&8.58$\,\pm\,$0.03&143$\,\pm\,$36&<9.7&<10.53&8.44$\,^{0.32}_{0.16}$\\
4497&2.44&8.62$\,\pm\,$0.04&161$\,\pm\,$63&<9.68&<10.44&8.49$\,^{0.36}_{0.2}$\\
5094&2.17&8.71$\,\pm\,$0.08&$<$136&9.88$\,\pm\,$0.11&10.75$\,\pm\,$0.11&$<$8.41\\
3324&2.31&8.81$\,\pm\,$0.04&162$\,\pm\,$60&9.93$\,\pm\,$0.12&10.64$\,\pm\,$0.12&8.49$\,^{0.34}_{0.19}$\\
24763&2.46&8.66$\,\pm\,$0.07&$<$136&9.41$\,\pm\,$0.2&10.36$\,\pm\,$0.2&$<$8.42\\
13296&2.17&8.82$\,\pm\,$0.05&216$\,\pm\,$66&9.54$\,\pm\,$0.08&10.23$\,\pm\,$0.08&8.61$\,^{0.31}_{0.17}$\\
\hline  
\hline
Low-metallicity stack$^{a}$ (3 galaxies) & 2.23 & 8.51 $\pm$ 0.07 &  105 $\pm$ 21 &  9.63 $\pm$ 0.13 & 10.83 $\pm$ 0.13 & 8.30$\,^{0.15}_{0.30}$\\
High-metallicity stack$^{b}$ (7 galaxies) & 2.26 & 8.72 $\pm$ 0.08 &  103 $\pm$ 21 &  9.67 $\pm$ 0.06 & 10.52 $\pm$ 0.06 & 8.29$\,^{0.15}_{0.30}$\\
\hline
\end{tabular}
Columns from left to right: 3D-HST v4 catalog ID, redshift, metallicity (\citealt{Bian2018} calibration), 1.2 mm ALMA band-6 flux, CO J$=$3--2 line luminosity, derived \h2 mass, and derived dust mass. The last two rows correspond to the properties of the stack of all galaxies in the sample and the CO J$=$3--2 non-detections. The redshift and metallicity for the stacks correspond to the mean value and standard deviation. $^a$12 + log(O/H)$<$8.6. $^b$12 + log(O/H)$>$8.6.
\end{table*}

\section{Data}
\label{sec:data}
\subsection{Sample and data}

The galaxies selected for the present work are from the  Cosmic Evolution Survey (COSMOS) field with available data from the MOSDEF survey \citep{Kriek2015}. As described in \citet{Kriek2015}, the Keck/MOSFIRE spectrograph \citep{McLean2012} was used to obtain rest-optical spectra for MOSDEF galaxies, covering the strong optical emission lines [OII]$\lambda \lambda$3726,3729, H$\beta$, [OIII]$\lambda \lambda$4959,5007, H$\alpha$, [NII]$\lambda$6584, and [SII]$\lambda \lambda$6716,6731. Such measurements enable the robust determination of the gas-phase metallicity of the targeted galaxies \citep{Sanders2018,Sanders2021}. From the MOSDEF sample, we selected galaxies that have ALMA band-6 1.2~mm coverage from programme 2019.1.01142.S (presented in \citealt{Shivaei2022}) and CO J$=$3--2 coverage from programme 2018.1.01128.S (presented in \citealt{Sanders2022}) that are not classified as mergers (via the visual classification of close pairs or disturbed systems using multi-wavelength rest-optical data; \citealt{Kartaltepe2012}). These criteria resulted in a sample of ten galaxies in the redshift range 2.1--2.5, with stellar masses ranging from  $10^{10.3}$ to $10^{10.6}\,\rm{M}_\odot$, star-formation rates (SFRs) ranging from 35 to 145 ${M}_\odot\,\rm{yr}^{-1}$ (see \citealt{Shivaei2022} for more information on the derivation of the stellar masses and SFRs), and metallicities in the range 8.4 < $12 + \log{(\rm{O/H})} < 8.8$ (0.5 -- 1.25 Z$_\odot$). A detailed description of the metallicity estimation is provided in Sect.~\ref{sec:metallicity} and an overview of the sample properties in Table~\ref{tab:sample}. 

The stellar mass, SFR, and metallicity distributions of our sample are presented in Fig.~\ref{fig:MS} and compared to galaxies at redshifts $z=1.5 - 2.5$ from the MOSDEF survey. Our sample of galaxies is representative of the diversity in metallicities of galaxies with stellar masses of $\sim 10^{10.5}\,\rm{M}_\odot$ and encompasses the upper envelope of the main sequence of star formation at this stellar mass and redshift (as derived in \citealt{Shivaei2015}).

A detailed description of the ALMA data calibration, imaging, and flux measurements is included in the works that present the band-6 and band-3 data \citep[respectively,][]{Shivaei2022, Sanders2022}. Six galaxies in our sample are detected at 1.2~mm  (\citealt{Shivaei2022}) with a signal-to-noise ratio (S/N) of at least 2.3, which is sufficient given the detailed prior knowledge of these galaxies.  The CO J$=$3--2 emission line is robustly detected with a S/N of at least 3.8 for six of the ten galaxies of our sample \citep{Sanders2022}. For another three of the galaxies, both the 1.2~mm and CO J$=$3--2 emission are detected. Finally, the remaining galaxy is detected neither at 1.2~mm nor in CO J$=$3--2. With this sample, we can robustly constrain the \dtg\ for three galaxies, while placing upper limits for three and lower limits for three. The band-6 1.2~mm flux densities and CO J$=$3--2 line luminosities of our sample of galaxies (as well as derived quantities) are tabulated in Table~\ref{tab:sample}. 

\begin{figure}
\centering
\includegraphics[width=\linewidth]{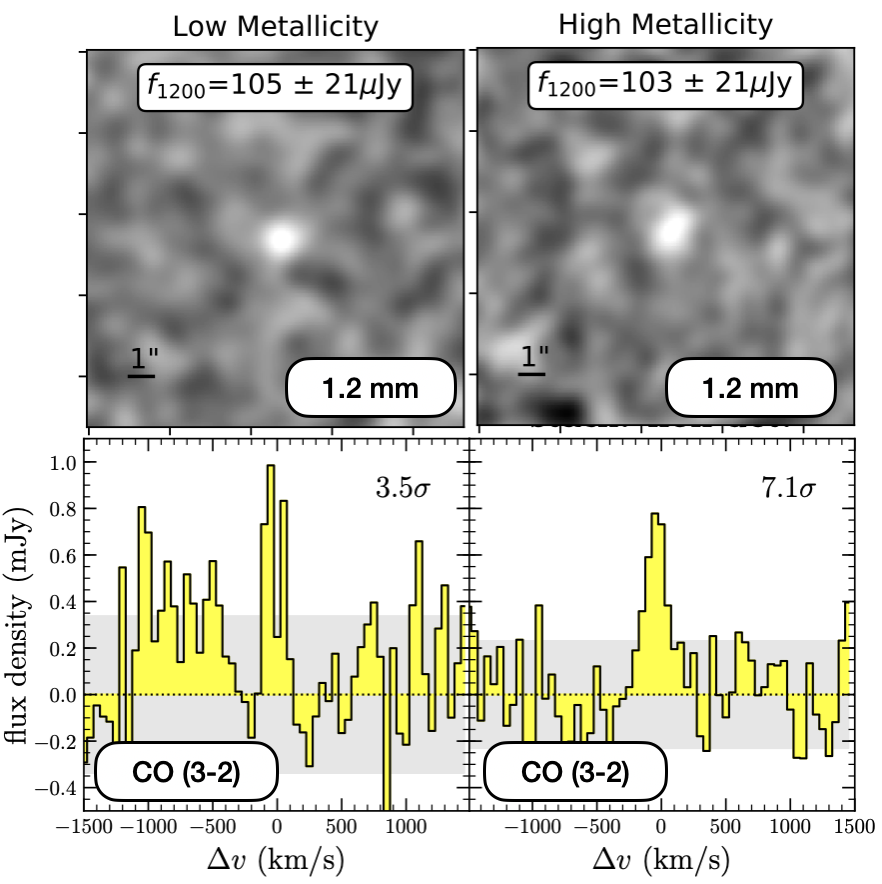}
\caption{ The top panels show the 1.2 mm dust-continuum detections of the stacked datasets corresponding to the low-metallicity galaxies (left; 12 + log(O/H) $<$8.6; three galaxies) and high-metallicity galaxies (right; 12 + log(O/H) $>$8.6; seven galaxies)  presented in this work. The bottom panels show the CO spectral profile for the same stacked galaxies. \label{fig:stacked_images} }
\end{figure}

We additionally used stacking techniques to include information from all the galaxies in our analysis. The adopted methodology is discussed in \citet{Shivaei2022} and \citet{Sanders2022}. We created stacked images and composite spectra for the low-metallicity galaxies (12 + log(O/H) $<$8.6; three galaxies)  and high-metallicity galaxies (12 + log(O/H) $>$ 8.6; seven galaxies) presented in this work. The CO J$=$3--2 and 1.2~mm emission is detected for both sets of stacked galaxies. Figure~\ref{fig:stacked_images} shows the dust-continuum detection (top row) and CO J$=$3--2 profile (bottom row) of the two stacks. The median galaxy properties and fluxes of the stacks are tabulated in Table~\ref{tab:sample}.

\begin{figure*}
\centering
\includegraphics[width=0.495\linewidth]{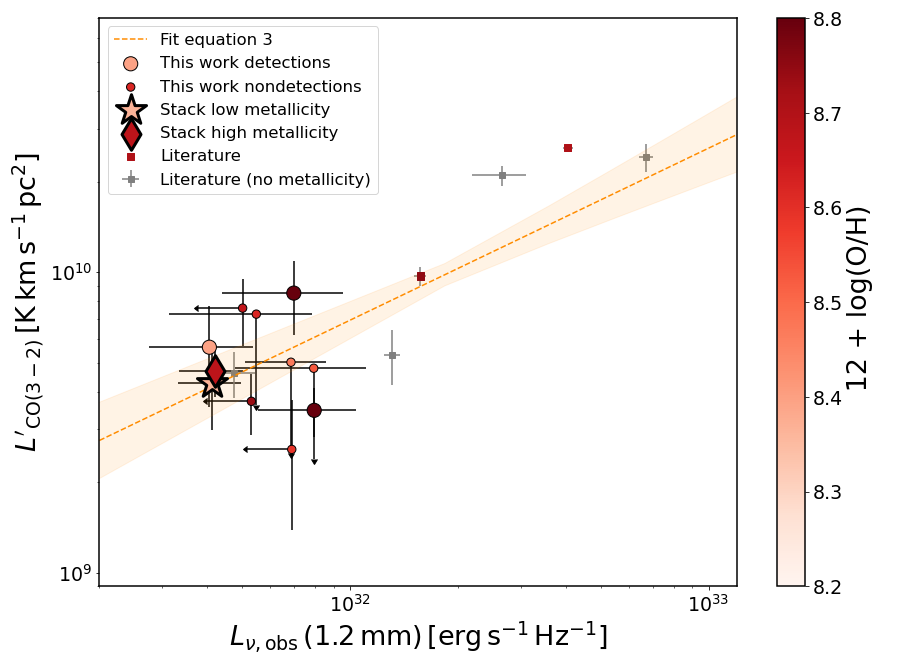}
\includegraphics[width=0.495\linewidth]{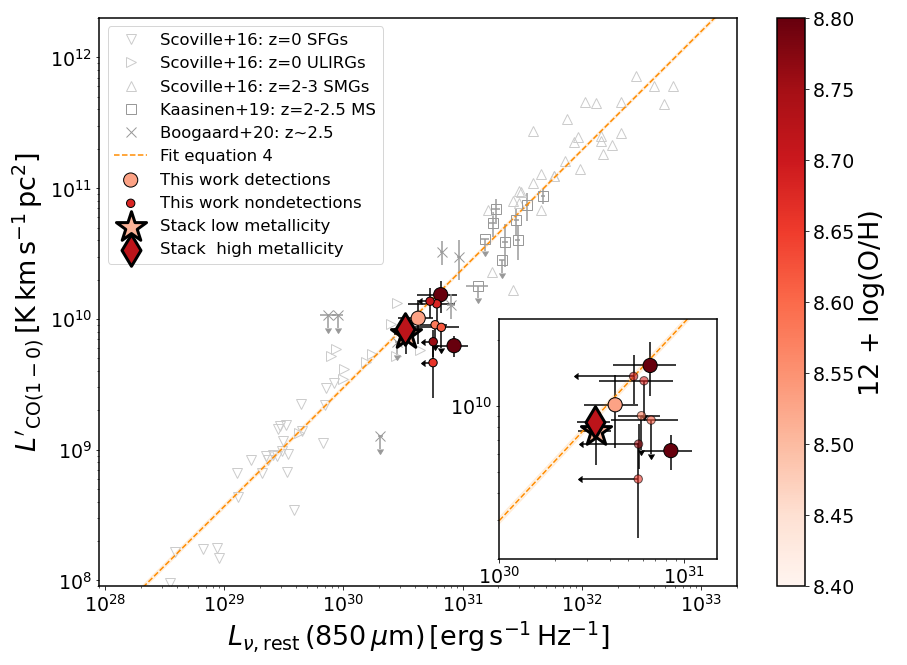}
\caption{Left: CO J$=$3--2 luminosity as a function of 1.2~mm luminosity for our sample of galaxies and the low- and high-metallicity stacks, compared with a compilation taken from \citet{Boogaard2020}, \citet{Popping2017}, and \citet{Talia2018}. The dashed orange line {and shaded area around it, indicating the one-sigma uncertainty,} correspond to the fit presented in Eq.~\ref{eq:fit1}. Right: CO J$=$1--0 luminosity of our sample of galaxies (assuming $r_{1-0/3-2} = 1.8$) as a function of the 850 \micron rest-frame luminosity (calculated assuming a thin grey body with $\beta = 1.5$ and $T_{\rm dust} = 25\,\rm{K}$), compared with a sample of local and high-redshift galaxies from \citet{scoville16} and $z=1-2$ galaxies from \citet{Kaasinen2019} and \citet{Boogaard2020}. The dashed orange line {and shaded area around it (note that the uncertainty range is very narrow)} correspond to the fit presented in Eq.~\ref{eq:fit2}. In both panels the galaxies are colour-coded based on their metallicity, when available. \label{fig:CO_vs_continuum} }
\end{figure*}

\subsection{Galaxy properties}
\subsubsection{Metallicity}
\label{sec:metallicity}

Gas-phase oxygen abundances (i.e. metallicities) were estimated using ratios of strong rest-optical emission lines.
The ten targets presented in this work have $\ge3\sigma$ detections of H$\beta$, [OIII]$\lambda$5007, H$\alpha$,
 and [NII]$\lambda$6584, so the O3N2=([OIII]/H$\beta$)/([NII]/H$\alpha$) indicator can be used.
Calibrations between O3N2 and metallicity have been constructed using observations of $z\sim0$ HII regions and galaxies \citep[e.g.][]{Pettini2004,Curti2020}.
However, local Universe metallicity calibrations are suspected to be inaccurate at high redshifts due to evolving ionisation conditions in HII regions \citep[e.g.][]{Steidel2016,Strom2018,Topping2020a,Sanders2020}.
Using direct-method metallicity measurements of galaxies at $z=1.7-3.6$, \citet{Sanders2020} find that the calibrations of \citet[B18]{Bian2018} constructed from extreme local galaxies selected to be analogues of $z\sim2$ galaxies appear to yield reliable metallicities for high-redshift samples (rather than using e.g. the \citealt[PP04,]{Pettini2004} calibration).
One caveat is that \citet{Sanders2020} could only test calibrations that use line ratios of [OII], H$\beta$, [OIII], and [NeIII] and could not test ratios involving [NII], such as N2=[NII]/H$\alpha$ and O3N2.
However, \citet{Sanders2022} re-normalised the \citet{Bian2018} O3N2 calibration such that it yields metallicities consistent with those based on [OII], H$\beta$, [OIII], and [NeIII].
For all $z>2$ sources in this work, we utilised this re-normalised O3N2 calibration from \citet{Sanders2022}:
$
12+\log(\mbox{O/H})=9.03-0.39\times \mbox{O3N2}.
$

\subsubsection{Dust mass}
\label{sec:dust}
We calculated the dust mass ($M_{\rm dust}$) of the galaxies from their dust-continuum emission at 1.2~mm following the approach outlined in \citet{Shivaei2022}. They find that, at an observed wavelength of 1.2\,mm, cold dust is predominantly responsible for the observed dust-continuum emission of sub-solar- and solar-metallicity main-sequence galaxies at $z=2.1-2.5$, making the 1.2\,mm emission a good dust mass diagnostic. We derived dust masses from the ALMA 1.2\,mm flux densities, assuming an optically thin modified black body, as\begin{equation}\label{eq:mdust}
    M_{\rm dust} = \frac{(S_{\nu}/f_{\rm CMB})~D_{\rm L}^2(z)}{B_{\nu}(T)~(1+z)~\kappa_{\nu}(\beta)},
\end{equation}
where $S_{\nu}$ is the 1.2\,mm flux density (in the observed frame), $f_{\rm CMB}$ is the correction factor for the cosmic microwave background (CMB) effect on the background at the redshift of the targets \citep[Eq. 18 in][]{dacunha13}, $D_{\rm L}(z)$ is the luminosity distance to redshift $z$, $B_{\nu}(T)$ is the Planck function with dust temperature $T$, and $\kappa_{\nu}(\beta)$ is the dust grain absorption cross-section per unit mass at frequency $\nu$ with a functional form of $\kappa_0(\frac{\nu}{\nu_0})^{\beta}$, where $\kappa_0$ is the opacity at $\nu_0$ and $\beta$ is the sub-millimetre emissivity index. Following \citet{Shivaei2022}, we accounted for the systematic uncertainty using a Monte Carlo sampling approach in which we drew dust temperatures from a Gaussian distribution with a mean of 25\,K and $\sigma=5$\,K and calculated the final dust mass and its uncertainty as the median and dispersion on the resulting dust masses.  We furthermore assumed an emissivity index of $\beta=1.5,$ and, based on this choice, an opacity of $\kappa_0=0.232$\,m$^2$\,kg$^{-1}$ at 250\,{\um} was adopted \citep{draine03,bianchi13}. We refer the reader to Appendix\,B of \citet{Shivaei2022} for a detailed discussion of the derivation of dust masses for the presented sample of galaxies and the systematic uncertainties.

\subsubsection{\h2 mass}
\label{sec:H2}
The \h2 masses of the galaxies in this paper were calculated from their CO J$=$3--2 luminosities. We first estimated the CO J$=$1--0 emission assuming a CO J$=$3--2 to J$=$1--0 excitation correction factor of $r_{1-0/3-2} = 1.8$ (\citealt{Tacconi2018}, appropriate for $z=$2.1-2.5 star-forming main-sequence galaxies). To calculate the \h2 mass of the galaxies, we adopted the \citet{Accurso2017} metallicity-dependent recipe for the CO-to-\h2 conversion factor, $\alpha_{\rm CO}$, where
\begin{equation}
    \alpha_{\rm CO} = 10^{14.752 - 1.623[12 + \log{\rm{(O/H)}}]},
\end{equation}
without taking the minimal dependence on the offset from the main sequence into account. \citet{Sanders2022} estimated $\alpha_{\rm CO}$ for a number of $z\sim 2$ galaxies (including the ones presented in this work) by calculating the \h2 mass from dynamical measurements and from the inverse Schmidt-Kennicutt relation, finding the \citet{Accurso2017} relation to be consistent with their results. We discuss the systematic uncertainties in the \h2 mass and their impact on our conclusions in more detail in Sect.~\ref{sec:discussion}.

\begin{figure*}
\centering
\includegraphics[width=0.75\linewidth]{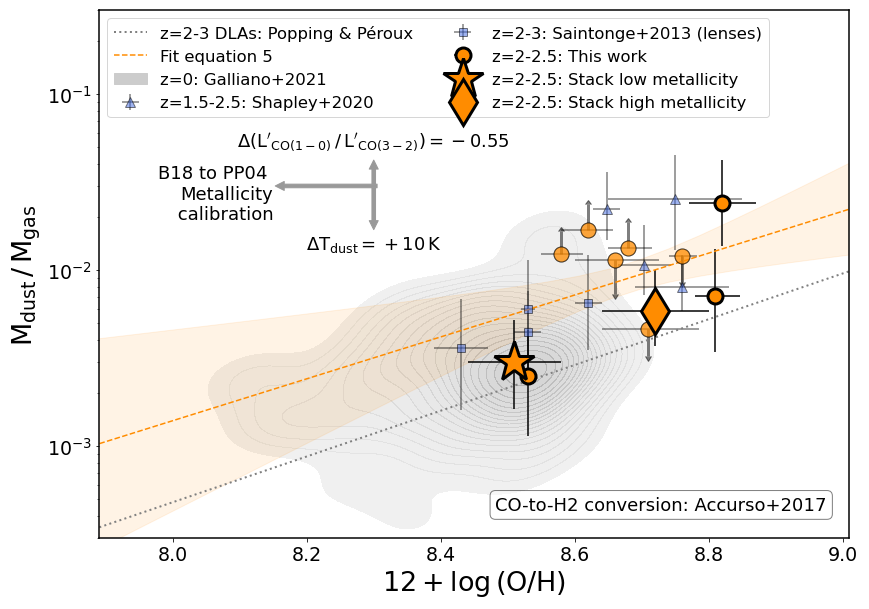}
\caption{\dtg\ of galaxies as a function of their metallicity. The orange circles correspond to the individual galaxies  presented in this work and the orange star and diamond  to the  stacks. Circles with upward or downward pointing arrows correspond to 2$\sigma$ limits for the non-detections. A $z>1.5$ literature compilation from \citet{Saintonge2013} and \citet{Shapley2020} is shown with blue squares and circles, respectively. The dashed orange line {and shaded area around it, indicating the one-sigma uncertainty,} correspond to the fit presented in Eq.~\ref{eq:DTG_fit}. {Overplotted is the best fit relation presented by \citet{Popping2022}, based on absorption studies of damped Lyman-$\alpha$ absorbers.} A $z=0$ compilation taken from \citet{Galliano2021} is plotted as a grey-shaded area, and we include the \citet{DeVis2019} fit to local luminous galaxies as a dashed black line. The arrows included in the figure mark the systematic uncertainty introduced when adopting other assumptions for the dust temperature, metallicity calibration, and CO excitation properties of the galaxies.    \label{fig:DTG} }
\end{figure*}

\section{Results}
\label{sec:results}
\subsection{The relation between CO and dust-continuum emission}
We present the CO J$=$3--2 luminosity of the galaxies as a function of their 1.2~mm dust continuum brightness in the left panel of Fig.~\ref{fig:CO_vs_continuum}, colour-coded by metallicity. For comparison, we include main-sequence galaxies at similar redshifts drawn from the literature where the same information is available (CO J$=$3--2 line emission and ALMA band-6 continuum). The galaxies presented in this work correspond to the faintest galaxies at $z\sim2$ for which both CO J$=$3--2 and dust continuum information are available. We find a strong correlation between the CO and dust continuum luminosities for the combined set of literature galaxies and the sample of galaxies presented in this work, anchored by the low-luminosity sample presented herein. The combined sample of galaxies presented in this work and the literature has a Pearson rank correlation coefficient of 0.89 and a p-value of $5\times10^{-6}$ and can be described as\begin{multline}
\label{eq:fit1}
\log{(L'_{\rm{CO(3-2)}})} = \\
(0.57 \pm 0.13) \times \log{(L_{\nu,\rm{obs}}\,(1.2\,\rm{mm}))} - (8.48 \pm 4.15).
\end{multline}
{This fitting relation was obtained via Monte Carlo sampling one thousand times over the uncertainty of the CO and 1.2~mm luminosity of every galaxy and applying a least-square fit to the individual samples. The final fit and uncertainty correspond to the median and one-sigma deviation of the individual fits.}

The relation presented in Eq.~\ref{eq:fit1} is sub-linear, {possibly} indicating the existence of a second-order trend between the 1.2~mm luminosity and CO J$=$3--2 luminosity of galaxies. We do not find an obvious second-order dependence of this correlation as a function of gas-phase metallicity. Galaxies with similar metallicities at a fixed 1.2~mm luminosity can be found both above and below the fit presented in Eq.~\ref{eq:fit1}. Figure~\ref{fig:CO_vs_continuum} also includes data points that correspond to stacks of the low- and high-metallicity galaxies. These stacks are consistent with the results based on the individual galaxies and do not show a clear trend with metallicity either. 

The right panel of Fig.~\ref{fig:CO_vs_continuum} shows the CO J$=$1--0 luminosity of the galaxies (derived assuming a CO J$=$1--0 to CO J$=$3--2 luminosity ratio of $r_{1-0/3-2}=$1.8, \citealt{Tacconi2018}) as a function of their rest-frame 850 \micron luminosity (assuming that the dust-continuum spectral energy distribution can be described as a grey body with $\beta=1.5$ and $T_{\rm dust}=25\,K$). For comparison, we also include local star-forming galaxies, ultra-luminous infrared galaxies, and the $z=2-3$ sub-millimetre galaxies presented in \citet{scoville16}.
We furthermore include $z\sim2$ main-sequence galaxies for which CO J$=$1--0 is available, taken from \citet{Kaasinen2019} and \citet{Boogaard2020}. 
We find that our sample of galaxies and their stacks connect the $z=0$ and $z>2$ galaxies observed to date, indicating a continuous relation between galaxy CO and dust-continuum luminosity, with a scatter similar to the scatter found for $z>2$ main-sequence and sub-millimetre galaxies. The relation between CO J$=$1--0 and rest-frame 850 \micron luminosity for the compilation of all the literature galaxies and the galaxies presented in this work can be described by a near-linear relation:
\begin{multline}
\label{eq:fit2}
\log{(L'_{\rm{CO(1-0)}})} = \\
(0.91 \pm 0.01) \times \log{(L_{\nu,\rm{rest}}\,(850\,\mu\rm{m}))} - (17.78 \pm 0.16).
\end{multline}
{The fitting relation was obtained by performing a least-squares fit to the data a thousand times, each time perturbing it by Gaussian-sampling over the uncertainty of each galaxy in CO and 850~\micron\ luminosity. The final fit and uncertainty were obtained by taking the median and one-sigma deviation of all individual fits (Fig.~\ref{fig:CO_vs_continuum})}

The near-linear shape of Eq.~\ref{eq:fit2} over four orders of magnitude indicates there is only {a} weak second-order dependence between the CO J$=$1--0 and rest-frame 850~\micron luminosity of galaxies studied so far (in contrast to the relation between CO J$=$3--2 and observed 1.2~mm luminosity). Indeed, we do not find a second-order trend as a function of gas-phase metallicity for the relation between galaxy CO J$=$1--0 and 850~\micron luminosity for the sample of galaxies presented herein, although more high-redshift measurements at sub-solar metallicities will be necessary to confirm this. 

\subsection{The dust-to-gas mass ratio of galaxies}
The unique combination of metallicity, dust-continuum, and CO information of the sample of galaxies presented in this work allows us to place additional constraints on the DTG of $z\sim 2$ sub-solar-metallicity luminous galaxies. In Fig.~\ref{fig:DTG} we show the \dtg\ of galaxies as a function of their metallicity. The \dtg\ is defined as $M_{\rm H2}/M_{\rm dust}$ and excludes the contribution from atomic hydrogen, which, according to simulation predictions, corresponds to $\sim25$\% of the total gas budget for galaxies with stellar masses of around $10^{10}\,\rm{M}_\odot$ (e.g. \citealt{Popping2015}).  We compared our sample of galaxies to a $z = 1.5 - 3$ literature compilation based on data presented in \citet{Saintonge2013} and \citet[with dust-continuum and CO information from \citealt{Dunlop2017}, \citealt{Aravena2020}, \citealt{Gonzalez2019}, and \citealt{Riechers2020}]{Shapley2020}, as well as the $z=0$ compilation presented in \citet{Galliano2021} and the fitting equation presented by \citealt{DeVis2019} for a sample of local luminous galaxies (though we note that the \citealt{DeVis2019} \dtg\ also takes \hi into account).\footnote{The \citet{Galliano2021} and \citet{DeVis2019} analyses employ metallicities derived with the \citet{Pilyugin2016} `S calibration', which uses a combination of the [O~\textsc{iii}]/H$\beta$, [N~\textsc{ii}]/H$\alpha$, and [S~\textsc{ii}/H$\alpha$] ratios.  Our adopted $z\sim2$ calibration uses only the [O~\textsc{iii}]/H$\beta$ and [N~\textsc{ii}]/H$\alpha$ ratios in O3N2 because [S~\textsc{ii}/H$\alpha$] was not included in the high-redshift calibration set of \citet{Bian2018}.  Despite the slight difference in the line ratio set used to derive metallicities for the local and $z\sim2$ samples, both calibrations are importantly based on direct-method $T_e$ metallicities, and as such our comparison avoids the offset of a factor of $\sim2$ in metallicity between $T_e$-based and photoionisation-model-based metallicities (e.g. \citealt{Kewley2008}, \citealt{Sanders2021}.} For consistency, we re-calculated the dust and gas mass for the galaxies presented in \citet{Shapley2020} and \citet{Saintonge2013} as outlined in Sects.~\ref{sec:dust} and \ref{sec:H2}. We only included the galaxies from the \citet{Galliano2021} sample for which CO observations are available and re-calculated the \dtg\ adopting the same CO-to-\h2 conversion factor as in Sect.~\ref{sec:H2}.

We find that the \dtg s\ of the presented galaxies are consistent with an increasing trend in the \dtg\ as a function of metallicity. This trend is also clearly seen when comparing the \dtg\ of the stacked set of low-metallicity  galaxies to the stack of the high-metallicity galaxies.
The \dtg\ of the detected galaxies is consistent with the \dtg\ obtained for the galaxies presented in \citet{Shapley2020}. The \dtg\ of the high-metallicity stack is $\sim$0.15 dex below the average \dtg\ of the galaxies presented in \citet{Shapley2020}. The \dtg\ found for the galaxies presented herein is consistent with the \dtg\ calculated for the galaxies presented in  \citet{Saintonge2013}. 

{The relation between the \dtg\ and metallicity of the $z\sim2-2.5$ galaxies (literature and this work) can be described as a linear function (accounting for the upper and lower limits):
\begin{multline}
\label{eq:DTG_fit}
\log{\rm{\dtg}} = \\
(1.25 \pm 0.75)[12 + \log{\rm{(O/H)}}] - (12.99 \pm 6.52).
\end{multline}
The fitting relation was obtained by performing a least-square fit to the data a thousand times, each time perturbing the data by Gaussian-sampling over the uncertainty in metallicity and the \dtg\ for each galaxy, while also accounting for upper and lower limits. The final fit and uncertainty were obtained by taking the median and one-sigma deviation of all individual fits.}

The \dtg\ of the $z=$2.1--2.5 MOSDEF galaxies, probing a metallicity range from 12 + log(O/H)$=$8.4--8.8, and their stacks are consistent with the \dtg\ of the $z=0$ compilation of galaxies from \citet{Galliano2021}, which probed a similar metallicity range. The data and stacks presented in Fig.~\ref{fig:DTG} are furthermore consistent with the fitting equation presented by \citet{DeVis2019} for a sample of local luminous galaxies. Although the number of $z=$2.1-2.5 galaxies for which the \dtg\ is constrained is low compared to $z=0$ (especially at sub-solar metallicities), this suggests that the relation between the \dtg\ and 12 + log(O/H) remains constant from $z=0$ to $z=2.5$ at metallicities larger than half solar. Further CO observations and multi-band dust-continuum observations of sub-solar-metallicity galaxies will be necessary to robustly confirm this result, extend it to lower metallicities, and quantify the scatter.

\section{Discussion}
\label{sec:discussion}
We have shown that the \dtg\ of galaxies at $z=$2.1--2.5 increases as a function of metallicity and that there is no evidence of an evolution in this relation from $z=0 $ to $z=2.5$ (Fig.~\ref{fig:DTG}). A lack of evolution had been reported earlier for $z=$1.5--2.5 galaxies in \citet{Shapley2020} and through studies of the dust-depletion patterns of damped Lyman-$\alpha$ absorbers (see the review by \citealt{Peroux2020} and more recently \citealt{Popping2022}). A similar conclusion can be drawn based on the sample of lensed galaxies presented in \citet{Saintonge2013}. The current work is the first to draw this conclusion based on a sample of un-lensed luminous $z\sim2$ main-sequence galaxies that includes galaxies with sub-solar metallicities, exploring a new regime in main-sequence parameter space and detection techniques. 

We acknowledge that the presented results and conclusions are dependent on a number of assumptions. For example, the slope of the relation between the \dtg\ and 12 + log(O/H) depends on the CO-to-\h2 conversion factor used. Nevertheless, both the conclusions that the \dtg\ of galaxies increases as a function of gas-phase metallicity and that this relation is constant with time remain true when choosing different CO-to-\h2 conversion factors (for example, the metallicity-dependent conversion factor relations presented in \citealt{Genzel2012}, \citealt{Schruba2012}, \citealt{Bolatto2013}, \citealt{Amorin2016}, and \citealt{Madden2020}). The latter conclusion remains true since the \dtg s\ of the $z=0$ and $z=$1.5-3 samples of galaxies change by the same amount (at fixed metallicity) when the CO-to-\h2 conversion factor changes, as long as the same relation is used at $z=0$ and $z\sim2.3$.\footnote{\citet{Sanders2022} demonstrated that the \citet{Accurso2017} relation agrees best with estimates of the $z\sim2$ $\alpha_{\rm CO}$--metallicity relation with \h2 masses derived from dynamical mass measurements or when using an inverse Schmidt-Kennicutt relation that links the SFR surface densities of galaxies to their \h2 surface densities.} { The former conclusion, an increase in \dtg\ with metallicity, remains true for all the abovementioned CO-to-\h2 conversion factors, though the slope of the relation increases strongly when adopting CO-to-\h2 recipes with a strong dependence on metallicity (e.g. \citealt{Schruba2012} and \citealt{Madden2020}).} 

The arrows in Fig.~\ref{fig:DTG} indicate the systematic uncertainty in the \dtg\ and gas-phase metallicity when assuming a different dust temperature, metallicity calibration, or CO excitation correction. These differences can be significant (up to 0.2 dex in metallicity or ~0.3 dex in \dtg) and change the intercept of the relation between the \dtg\ and 12 + log(O/H). It is likely that the differences could change the slope of this relation since, for example, the CO excitation conditions or dust temperature of sub-solar-metallicity galaxies may be different from those of solar-metallicity galaxies. It is, however, unlikely that these uncertainties change the slope of the relation to such an extent that the \dtg\ of galaxies decreases with increasing metallicity. 

{An additional point of uncertainty is the contribution of atomic gas to the \dtg\ from $z\sim2$ galaxies. In this work we have focused on CO emission as a tracer of the molecular gas budget of galaxies, potentially missing a significant contribution from HI. The missing contribution of HI may be (one of) the reason(s) for the offset with the metallicity -- the \dtg\ relation found by \citet{Popping2022} based on absorption studies (which are sensitive to the HI budget of absorbers) and the relations presented herein. Observations of local galaxies indeed demonstrate that the ratio between atomic and molecular hydrogen changes as a function of stellar mass and metallicity (see \citealt{Saintonge2022} for a recent review). 

While observations of HI in emission at $z\sim2$ are currently not yet available, we can resort to simulations to provide guidance. Semi-analytic and semi-empirical models predict a typical contribution by HI to the total cold gas budget of galaxies with stellar masses of $\sim10^{10}\,\rm{M}_\odot$ at $z\sim2$ of the order of 25\% \citep{Lagos2011, Popping2014, Popping2015}. More recent predictions of the HI versus H2 content of $z>1$ galaxies, obtained by extrapolating gas-scaling relations of local galaxies, suggest HI contributes to approximately 50\% or less of the cold gas budget of galaxies with $\sim10^{10}\,\rm{M}_\odot$ at $z\sim1$ \citep{Zhang2021}. Extrapolating the predictions by \citet{Zhang2021} to $z\sim2,$ we can expect an HI fraction of $\sim30$\% or less for this model. This is in contrast with recent predictions by \citet{Morselli2021}, who find an HI fraction well above 50\% for $z\sim2$ galaxies. This model is based on $z=0$ empirical relations between $M_{\rm H2}/M_{\rm HI}$ and either stellar mass or total gas surface density. It is unclear if these empirical trends also hold at $z>0$ (see e.g. changes in the predictions for some of these relations as a function of time presented in \citealt{Popping2015}). \citet{Chowdhury2020} used stacking techniques to obtain a measurement of the HI mass in emission of star-forming galaxies at $z=1$. Based on their findings, typical HI fractions significantly lower than 50\% can be derived at $z=1$. Based on the above works, it is not unlikely that the \dtg\ of the galaxies presented in this work is actually 30\% (or more) lower, which would bring it into better agreement with the findings by \citealt{Popping2022}. The HI fraction may also depend on stellar mass or metallicity, possibly causing a change in the slope of the relation. In the next decade, the Square Kilometre Array may be in a position to address this uncertainty in detail.}

{In this work we describe the relation between metallicity and the \dtg\ with a single power-law relation (Eq. \ref{eq:DTG_fit}). The metallicity range probed in this work is too narrow to rule out any non-linear trends. \citet{Remy-Ruyer2014} find that the relation between metallicity and the \dtg\ for galaxies in the local Universe (covering 2 dex in metallicity) can best be described by a double power law, with a break at $12 + \log{\rm{(O/H)}} = 8.1$. This is in contrast with results by \citet{DeVis2019}, who find that the metallicity--\dtg\ relation for local galaxies covering a similar metallicity range can be described by a single power law. Similarly, \citet{Popping2022} find that the metallicity--\dtg\ relation obtained for $z\sim2$ absorbers can be well described by a single power law covering a range of 2 dex in metallicity (with an increasing scatter at the low-metallicity end). Combined, these findings suggest that a non-linear trend between metallicity and the \dtg\ for $z\sim2$ galaxies in emission cannot be ruled out. Larger samples of sub-solar-metallicity galaxies with available CO and dust-continuum emission measurements will be necessary to test this.}

We have demonstrated that the relation between the CO J$=$3--2 (and derived J$=$1--0) luminosity and the 1.2 mm (and 850~\micron) luminosity of galaxies extends over multiple orders of magnitude (Fig.~\ref{fig:CO_vs_continuum}). The sample presented in this work extends the probed range in CO and dust continuum luminosities at $z\sim$2.1--2.5 to less massive galaxies (compared to \citealt{Kaasinen2019} and \citealt{Boogaard2020}). Additionally, we show that there is no second-order trend in this relation driven by the gas-phase metallicity of galaxies for the sample of galaxies presented in this work. This result indicates that the metallicity dependence of the \dtg\ for the presented sample of galaxies is primarily driven by the metallicity dependence of the CO--to--\h2 conversion factor and not by a metallicity dependence of the ratio between CO and dust-continuum luminosity.

The conclusions presented in this paper have a number of implications for our theoretical understanding of dust growth and destruction in galaxies. The non-evolution of the \dtg--metallicity relation indicates that the timescales for the balance of dust formation and destruction evolve little from $z=0$ to $z=2.5$. Over the last decade, various cosmological  hydro-dynamical and semi-analytical simulations have included the tracking of dust formation and destruction over cosmic time \citep{Popping2017,McKinnon2018, Hou2019,Li2019,Vijayan2019,Triani2020}. The constant normalisation of the relation between the \dtg\ and metallicity is in agreement with the simulations by \citet{Popping2017}, \citet{Li2019}, \citet{Vijayan2019}, and \citet{Triani2020} (see also the discussions in \citealt{Shapley2020} and \citealt{Popping2022}). The models by \citet{McKinnon2018} and \citet{Hou2019}, on the other hand, predict a measurable evolution from $z=2$ to $z=0$ (up to a factor of $\sim$3 for the \citealt{Hou2019} model), in disagreement with our results. None of the models predict a strictly linear relation similar to the fit presented in \citet{DeVis2019}, nor to the fits presented in \citet{Popping2022}, which are based on \dtg s inferred from absorption studies. Additional observational constraints on the \dtg\ of luminous galaxies at sub-solar metallicities will be necessary to draw more robust constraints.

Our conclusions also provide relevant insights for observational studies. Over the last few years, the dust-continuum emission of galaxies has become an efficient and widely used approach for estimating galaxy gas mass; this approach implicitly depends on the \dtg\  (\citealt{scoville16, Schinnerer2016, Kaasinen2019}). Our results imply that robust dust-based gas mass estimates of galaxies inferred
from millimetre dust-continuum observations must take gas metallicity into account (see also \citealt{Bertemes2018} and \citealt{Popping2022}). This effect will be especially relevant for metal-poor galaxies, for example low-mass galaxies and galaxies in the early Universe; in these cases, detecting classical gas-mass tracers, such as low-J transition CO and [CI] emission, is time-intensive, or currently not possible due to instrumental limitations.

The presented results provide an improved study of the functional shape and evolution of the \dtg--metallicity relation of galaxies over cosmic time. Better tracing of the functional shape, scatter, and evolution of the \dtg--metallicity relation will require better sampling of sub-solar-metallicity galaxies at various redshifts and improved constraints on dust temperature, CO excitation, and the CO--to--\h2 conversion factor for more reliable \h2 and dust masses. This will require a dedicated CO and multi-band dust-continuum follow-up of large samples of galaxies at various redshifts for which robust metallicities are available, such as those based on rest-frame optical nebular lines.

\begin{acknowledgements}
    We thank Leindert Boogaard, Melanie Kaasinen and Bahram Mobasher for useful discussions and feedback. We thank the referee for constructive comments. This paper makes use of the following ALMA data: ADS/JAO.ALMA\#2019.1.01142.S and ADS/JAO.ALMA\#2018.1.01128.S. ALMA is a partnership of ESO (representing its member states), NSF (USA) and NINS (Japan), together with NRC (Canada), MOST and ASIAA (Taiwan), and KASI (Republic of Korea), in cooperation with the Republic of Chile. The Joint ALMA Observatory is operated by ESO, AUI/NRAO and NAOJ. We acknowledge support from NSF AAG grants AST-1312780, 1312547, 1312764, 1313171, 2009313, and 2009085, grant AR-13907 from the Space Telescope Science Institute, grant NNX16AF54G from the NASA ADAP program. Support for this work was also provided through the NASA Hubble Fellowship grant \#HST-HF2-51469.001-A awarded by the Space Telescope Science Institute, which is operated by the Association of Universities for Research in Astronomy, Incorporated, under NASA contract NAS5-26555.

\end{acknowledgements}

%
%
\bibliographystyle{aa}
\bibliography{main.bib}




\end{document}